# IKLAX : A NEW MUSICAL AUDIO FORMAT FOR INTERACTIVE MUSIC

*Fabien Gallot*

iKlax Media
Biarritz,
France
www.iklax.com
fabien@iklax.com

*Owen Lagadec*

iKlax Media
Biarritz,
France
www.iklax.com
owen@iklax.com

*Myriam Desainte-Catherine*

SCRIME /
LaBRI - CNRS
Univ. of Bordeaux 1,
France
myriam@labri.fr

*Sylvain Marchand*

SCRIME /
LaBRI - CNRS
Univ. of Bordeaux 1,
France
sylvain.marchand
@labri.fr

## ABSTRACT

In this paper, we are presenting a new model for interactive music. Unlike most interactive systems, our model is based on file organization, but does not require digital audio treatments. This model includes a definition of a constraints system and its solver.

The products of this project are intended for the general public, inexperienced users, as well as professional musicians, and will be distributed commercially. We are here presenting three products of this project.

The difficulty of this project is to design a technology and software products for interactive music which must be easy to use by the general public and by professional composers.

## 1.INTRODUCTION

iKlax is an innovative music project for the development of active listening. Active listening is the possibility for the listener to change sound parameters during the listening of music.
Today, interactivity with music for the general public is limited to volume level variation, balance, and sometimes equalization. The iKlax Project wants to go beyond these limits, and to give the listener the ability to choose the instruments and/or voices he/she wants, the level variations for each element, and their location in space. Also, iKlax offers a constraints system on interactive parameters that allows the composer to define the amount of freedom available in his/her music.

There are quite a lot of scientific researches about interactive music, such as Pachet's works [1] and Camurri's works [2]. Like the *Semantic HIFI Project* of *IRCAM* [3], these works are often concerning digital audio treatments. For example, works about sound sources separation (like the extraction of instruments from a polyphonic music), do not provide acceptable results for musician ears yet. The implementation of these methods is complex and requires heavy digital audio treatments.

Contrary to these works, iKlax proposes a new music organization: this project does not affect the sound signal, but the music file organization. We present here, a musical file format with separated audio tracks. These tracks are broadcasted with sound parameters which are been subdued by a musical constraints system.
Also, instead of mixing all the tracks in the recording studio into one channel (or two for stereo), iKlax can preserve all tracks in a single file, and adds some sound parameters to render the music according to the composer's thoughts.

Adding in a new audio file organization, constraints on audio parameters imply elaborated internal data structures, which have been modelled through constraints graphs. A specific constraints solver had to be implemented.

The iKlax technology has been developed thanks to the iKlax Media company (Biarritz, France) and LaBRI (Bordeaux, France) researchers collaboration.

In section 2, we will present the iKlax technology and its products: the iKlax format, its specific functionalities, the iKlax Player – an interactive media player, and the iKlax Creator – the iKlax file creation software. We will define, in section 3, musical constraints on track selection. Then, in section 4, some technical details about the iKlax technology and its software products will be exposed. Finally, section 5 will present outlooks on further improvements.

## 2. GENERAL PRESENTATION OF THE IKLAX TECHNOLOGY

The iKlax project consists of an audio file format and two software products: a music player for listeners and a music editor software for composers.

### 2.1. iKlax Format

The *iKlax format* is a media container which gathers all the tracks of a musical piece and metadata which define constraints and reading parameters of the tracks.

Each recorded track in iKlax format has been stored separately from the other tracks, without cross compression on sound signal (specific multichannel compression). This process allows a better quality for sound restitution.

Two levels of interactivity have been implemented: track selection and mixing. Other will be developed in the near future as equalization or spatialization. Each one has its specific constraints. Only the composer can set constraints, if he/she wants. The listener cannot get around these artistic constraints.

The first level of interactivity is the track selection: the listener can choose the tracks he/she wants to listen to. The defined constraints are *exclusion*, *implication,* and a *group constraint*. A *group constraint* is a group of tracks or sub-groups with two constraints: a minimum and a maximum number of tracks which can be listened to, simultaneously. All constraints can be applied on tracks or on other constraints (for more details on these constraints, see section 3).

The second level of interactivity is mixing: the listener can change track levels. The process is on every track, or track groups (which coincide or not with selected constraints), the composer can set a maximum and a minimum level.

The other mixing constraints are: *equality*, i*nequality,* and *balance*. The *equality* constraint means that some elements have an equal level. The inequality constraint allows to keep some elements are at a higher level than other elements. The balance constraint is a group of elements with a constant level sum: in this group, if the level of an element decreases, the levels of the other elements increase.

These two levels of interactivity are already implemented. Other levels, such as equalization or spatialization, are still under development.

### 2.2. iKlax Player

The *iKlax Player* is the interactive player which allows listeners to discover all the composer's creativity on a music.

It is a classic media player that includes a new graphic zone. The right zone of figure 1 displays all tracks and track groups which have been defined by the composer. With a simple click, the user can select various versions of an instrument and listen to the result in real time.

The iKlax Player also has all the functionalities of a classic player.

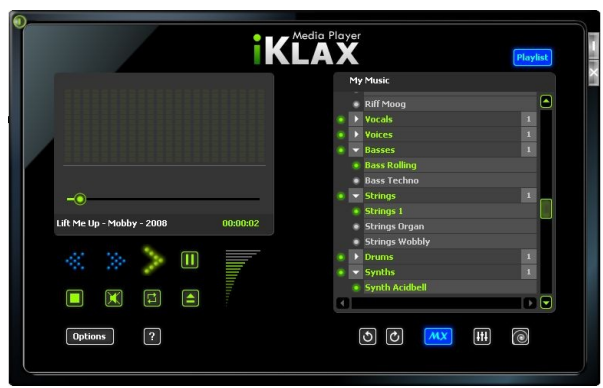

**Figure 1**. A screenshot of iKlax Player.

The exploration of a file and the selection of tracks to listen to are very easy. Tracks are stored in sub-groups that have been defined by the composer. These groups are displayed in a tree form (see figure 2), a common way of representation, well known by most people: all users (inexperienced users and professional musicians) that test the iKlax Player learn fast how it works.

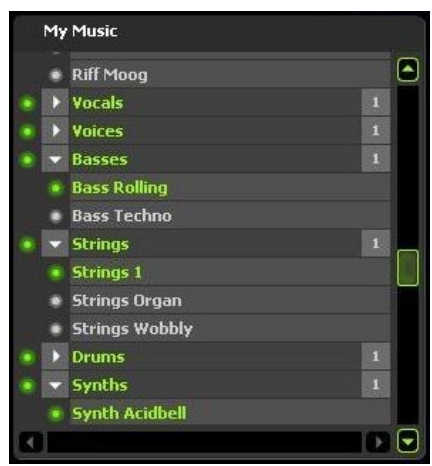

**Figure 2**. The tracks selection tree of iKlax Player.

## 2.3. iKlax Creator

The iKlax Creator (see figure 3) is the specific software product of the iKlax music creation suite. With its graphic user interface created by musicians for musicians, the iKlax creator permits all creative audacities, without changing traditional music recording. A constraints system (see section 2.1) gives the composer the possibility to control what the user can do or not, in respect of his/her creation. With the iKlax Creator, the composer can, in few clicks, create track groups and set constraints on his/her groups.

The iKlax Creator is the last stage in the music creation process, only replacing the final arrangement of the music. Today, the composer mixes all his tracks in a file. Now, with iKlax, the composer can export all his tracks to the iKlax Creator, to include constraints and create an iKlax file.

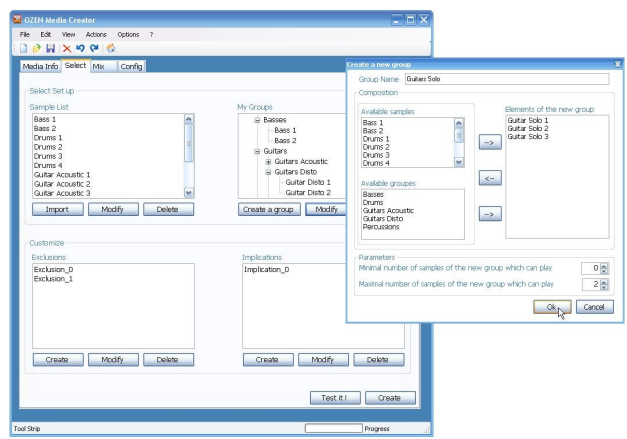

**Figure 3.** A screenshot of iKlax Creator.

## 3. SELECTION CONSTRAINTS DEFINITIONS

The following is an example of a file organization with selection constraint to understand the process of an iKlax file.

Let us consider a music with three instruments: a guitar, a piano, and drums. Each instrument has three different versions. Schematically, we have:
- Guitar (G) : G1, G2, G3
- Piano (P) : P1, P2, P3
- Drums (D) : D1, D2, D3

If no constraint has been defined by the artist, the user can make all combination he/she wants: G1, P1, D1 or G2, P1, D2… but also P1, B1 or G2, B3 or just P2…

**Exclusion**
If the artist thinks the version 1 of the guitar (G1) does not ring with the version 2 of the piano (P2), he/she may exclude the possibility of reading the two tracks simultaneously. He/She will then set an *exclusion* constraint.
With this constraint, if the listener plays the G1 track, the P2 track will automatically stop.

**Implication**
Conversely, the artist may want that version 2 of the guitar (G2) has to be played with version 1 of drums (D1). He/She will then set an *implication* constraint, which forces the reading of track D1, if the track G2 is being read.

**Group**
For artistic reasons, the artist may wish that at least one version of drums must been playing, but not the three of them at the same time. He can then create a group called "Drums" containing D1, D2, and D3 tracks, and set this group with a minimum of simultaneous playing tracks of one and a maximum of two. Thanks to this constraint, there will be always one or two drums playing at a time.

## 4. TECHNICAL CONSIDERATIONS

**Constraints Solver**
The constraints model on sound parameters and music organization have led to constraints graphs which may have cycles. So we choose a constraints solver on finite domains: *Gecode* [4].

The constraints solver has been designed to the point that the listener's action is the priority and the found solution is the nearest possible initialized state. This means that the listener's action may generate some automatic actions to preserve the music in a state which respects all composer constraints.
Sometimes, some actions may not be permitted: the listener will not be able do it. For example (with the music file of section 3), if the listener wants to stop all drums, but the composer has set a group of constraints on drums which forces at least one version of drums to be read. Then, the action to stop all drums will not work.

This solver process, with the search of the nearest initialized state solution, is based on the process used to solve problems of time constraints in *BOXES* [5-6]. The most significant difference between these two systems is that BOXES is modelled with the *Cassowary* library [7]. This library implements an optimization based on the simplex, contrary to *Gecode* – a library of resolution on finite domains.

**The core library**
The core library (without graphic elements) for the iKlax format has been implemented in C99 language, except the constraints solver which has been written in C++, because *Gecode* is a C++ library. All iKlax native libraries are portable.

**Graphic Users Interface**
Graphic users interfaces (GUIs) have been implemented for Microsoft Windows only, in a first time. However, GUIs for Mac OS X and Linux systems are under development.
The iKlax Player and Creator are now ready to be distributed.

## 5.PROSPECTS

The current and future work on the iKlax Project will concentrate on interactivity. At the time we are writing this paper, tracks selection and mixing have been implemented. The next works will be equalization and spatialization.

A problem we are not dealing with in this paper, but which must be rapidly solved, is the collision between constraints of different levels of interactivity.
For example, a composer can forbid to stop a track, but if he/she does not set a constraint on the sound level of this same track, the listener could simply decrease the sound level of the track down to zero. The track will be inaudible just as it was stopped.

This is the result of the constraints solvers independence on different levels of interactivity. At this time, there are no communications between solvers. Numerous collisions, sometimes very subtle, are identified between tracks selection, mixing, equalization, and spatialization.
In upcoming work, we must identify all *crossed constraints* and define processes or tools to help composers to set constraints. We ought to make a new specific constraints solver for interactive music with crossed constraints.

## 6.CONCLUSION

The iKlax project led to a new musical audio format, a new media player for interactive music, and a creation software for iKlax files. The format and software products are ready to be distributed.
The interactive music by music file organization, which is proposed in this paper, provides high sound quality and is an easy solution for the public active use.
This format has been created by and for musicians, in order to be adapted by them without difficulty and without a particularly high knowledge in computer science.
Thanks to the constraints system, artists have a total control on their music, while enabling active listening.

## 7.REFERENCES